\documentclass[pre,reprint,aps,raggedbottom]{revtex4-2}
\usepackage{ amssymb }
\usepackage{physics}
\usepackage{float}
\usepackage{comment}
\usepackage{amsmath}
\usepackage{graphicx}

\expandafter\def\expandafter\UrlBreaks\expandafter{\UrlBreaks%
  \do\/\do\-\do\.\do\:\do\=\do\_\do\?\do\0\do\1\do\2\do\3\do\4\do\5\do\6\do\7\do\8\do\9}

\newcommand{\conditionalEq}[2]{%
  \ifdim\columnwidth<\textwidth
    #1
  \else
    #2
  \fi
}

\newcommand{\zbar}{%
  \text{\ooalign{\hidewidth -\kern-.3em-\hidewidth\cr$z$\cr}}%
}

\usepackage{xurl}
\usepackage[colorlinks=true, urlcolor=blue, linkcolor=blue, citecolor=blue]{hyperref}

\begin{document}

\title{The Path Integral Monte Carlo Sign Problem Is Not Always NP-Hard: Harmonic Fermions Can Be Solved in Quadratic Time}

\author{A. Chaudhary$^1$}
\email{aarifchaudharyg@gmail.com}
\author{J. Valenzuela$^2$}
\email{jonasvalt@tamu.edu}
\author{Siu A. Chin$^2$}
\email{chin@physics.tamu.edu}
\affiliation{$^1$Hendrix Industries, Sealy, Texas 77474, USA}
\affiliation{$^2$Department of Physics and Astronomy, Texas A\&M University, College Station, TX 77843, USA}

\begin{abstract}
In the Path Integral Monte Carlo (PIMC) simulation of fermions in a harmonic trap, with and without pairwise harmonic interactions, the partition functions for any discrete number of imaginary time slices (or beads) and for any choice of the short-time propagator can be analytically obtained from the contracted determinant form of the propagator. This work shows that the resulting recursion relation can be reformulated in the $\lambda$-ring language, yielding a closed-form finite-bead partition function in two dimensions in terms of permutation statistics. This closed-form partition function can be evaluated by a special algorithm in $O(n^2)$ time, providing an exact and numerically stable scheme for reproducing the energies of the original (undoable) fermion PIMC simulation for $n=10^4$ or more fermions. This result provides a concrete framework in which the numerical instability of the sign problem is completely bypassed, and serves as a counterexample to the prevailing view that all truly fermionic PIMC sign problems are NP-hard.

\end{abstract}

\maketitle

\section{Introduction}
The fermionic sign problem remains a central obstacle in quantum many-body simulations, arising from the use of a determinantal free-fermion propagator, or equivalently, from
cancellations between different permutation sectors of the path integral. In Path Integral Monte Carlo (PIMC), these cancellations lead to an exponential degradation of the signal-to-noise ratio\cite{tro05}, severely limiting accessible system sizes and temperatures. While the sign problem is well understood to originate from fermionic antisymmetry, explicit analytical control over its structure, particularly at finite imaginary time, remains elusive.

This work addresses the case of harmonic fermions in two dimensions, where the severe sign problem is known to limit quantum-dot calculations to $\lesssim 100$ electrons\cite{nor23,chin-exactPIMC}. For two-dimensional harmonic fermions, with or without pairwise harmonic interactions, the discrete path integral can be exactly integrated\cite{chin-exactPIMC} to yield a recursion relation\cite{NewtonIden} for the $n$-fermion partition function at any discrete number of imaginary time slices (or beads) and for any choice of the short-time propagator. Although the discrete path integrals are evaluated exactly, the resulting recursion is infected with an even more virulent form of the sign problem (see Conclusion) through alternating contributions that lead to catastrophic cancellations. However, we found that this recursive structure admits an exact algebraic reformulation through the $\lambda$-ring structure, yielding a closed-form partition function informed by permutation statistics and spectral degeneracy. Crucially, this
enables a numerically stable, subtraction-free evaluation of the partition function in $O(n^2)$ time. The sign problem is thus not eliminated but bypassed through algebraic reformulation: the same finite-bead fermion PIMC partition function is exactly reproduced, without assuming the energy spectrum, admitting a bosonic reformulation\cite{tro05}, or replacing the original PIMC problem by another method or approximation.

\section{The discrete imaginary time partition function}
\label{shp}

Consider the $d=1$ harmonic oscillator Hamiltonian
\begin{equation}
    \hat{H} = -\frac{1}{2}\frac{d^2}{dx^2} + \frac{1}{2}x^2 = \hat{T} + \hat{V}.
\end{equation}
The 1-bead short-time approximation to the imaginary-time propagator\cite{10.1063/5.0164086} $G_1(x', x, \epsilon) = \langle x' | e^{-\epsilon(\hat{T} + \hat{V})} | x \rangle$ is 
\begin{equation}
   G_1 = \frac{1}{\sqrt{2\pi\kappa_1(\epsilon)}} e^{-\mu_1(\epsilon)\frac{1}{2}x'^2} e^{-\frac{1}{2\kappa_1(\epsilon)}(x' - x)^2} e^{-\mu_1(\epsilon)\frac{1}{2}x^2}.
\label{g1}
\end{equation}
For the second-order PA propagator, $\kappa_1(\epsilon) = \epsilon$ and $\mu_1(\epsilon) = \epsilon/2$.
The $N$-bead propagator is then given by
\begin{equation}
    G_N
    = \frac{1}{\sqrt{2\pi\kappa_N(\tau)}}e^{-\frac{1}{2}\mu_N(\tau)(x'^2+x^2)}e^{-\frac{1}{2\kappa_N(\tau)}(x'-x)^2},
\end{equation}
where $\tau = N\epsilon$, and $\kappa_N(\tau)$ and $\mu_N(\tau)$ are coefficients\cite{10.1063/5.0164086} obtained after contracting $N$ short-time propagators of (\ref{g1}). The contracted $N$-bead propagator for $n$ fermions $G^N_n(x',x;\tau)$\cite{chin-exactPIMC}, where $x = (x_1, \dots, x_n)$ and likewise for $x'$, gives the corresponding $n$-fermion partition function
\begin{equation}
   Z_n^N = \frac{1}{n!}\frac{1}{\sqrt{(2\pi\kappa_N)^n}}\prod_{i=1}^{n} \int dx_i e^{-\mu_N(x_i^2)} \det(K),
\label{nferz}
\end{equation}
where $K_{ij}(x) = e^{-\frac{1}{2\kappa_N(\tau)}(x_i - x_j)^2}$.

By expanding out the determinant $\det(K)$, the $n$-fermion partition function has been
shown\cite{NewtonIden} to be determined recursively via
\begin{equation}
\begin{split}
\label{mainRecursion}
    Z_n^{N,d} &= \frac{1}{n} \sum_{i=1}^n (-1)^{i-1} (z_i^N)^d Z_{n-i}^{N,d},\\
    (z_n^N)^d &= \left(\frac{b^{n/2}}{1-b^{n}}\right)^d,
\end{split}
\end{equation}
where $b=\exp(-Nu)$, and $u$ is the portal parameter\cite{10.1063/5.0164086} defined by $\cosh(u)=\zeta_1=1+\kappa_1\mu_1$.

These results can also be applied to the case of pairwise harmonically interacting fermions in a two-dimensional harmonic trap,
\begin{equation}
\begin{split}
    H &= \frac{1}{2m} \sum_{i=1}^{n} \mathbf{p}_i^2 + \frac{1}{2} \sum_{i=1}^{n} \mathbf{r}_i^2 + \frac{1}{2}\lambda \sum_{i=1}^{n} \,\sum_{j=1}^{n} (\mathbf{r}_i - \mathbf{r}_j)^2,
\end{split}
\end{equation}
corresponding\cite{PhysRevE.57.3871,chin-exactPIMC} to one center-of-mass Hamiltonian with $\omega=1$ and a Hamiltonian for $(n-1)$ non-interacting fermions with $\omega = \sqrt{1+2n\lambda}$.
The partition function for the interacting case is then given in terms of the harmonic oscillator partition function by\cite{PhysRevE.57.3871}
\begin{equation}
\label{Zharmonic}
    Z_n = \frac{Z_1}{Z^*_1}Z_n^*,
\end{equation}
where $Z^*_1$ and $Z_n^*$ are defined by the changes $\mu_N \rightarrow \mu_N^* = \omega^2 \mu_N$ and $b \rightarrow b^* = e^{-Nu^*}$.

\section{Sign Problem in the Recursive Relation}
Although the discrete path integrals have been evaluated analytically, the resulting recursion (\ref{mainRecursion}) retains a variant form of the sign problem. The alternating terms in the recursion 
largely cancel one another, and the sum quickly vanishes beyond conventional machine precision. As we will show, a direct numerical evaluation of (\ref{mainRecursion}) is not possible unless exceedingly high numerical precision is used.

The required precision can be quantified by comparing the largest term in the recursion, which is roughly the bosonic partition function, to the smallest term, which is the fermionic partition function  (see Appendix~\ref{appx:precision} for details). The minimum number of bits of precision required to resolve their difference, $B_d(n,\tau)$, is approximately
\begin{equation}
    B_d(n,\tau) \approx \frac{\tau}{\ln(2)} (E_F^d - E_B^d),
    \label{eq:PrecisionReq}
\end{equation}
where $E_F^d$ and $E_B^d$ are the fermion and boson energies from their respective partition functions.

\begin{figure}[H]
    \includegraphics[width=\linewidth]{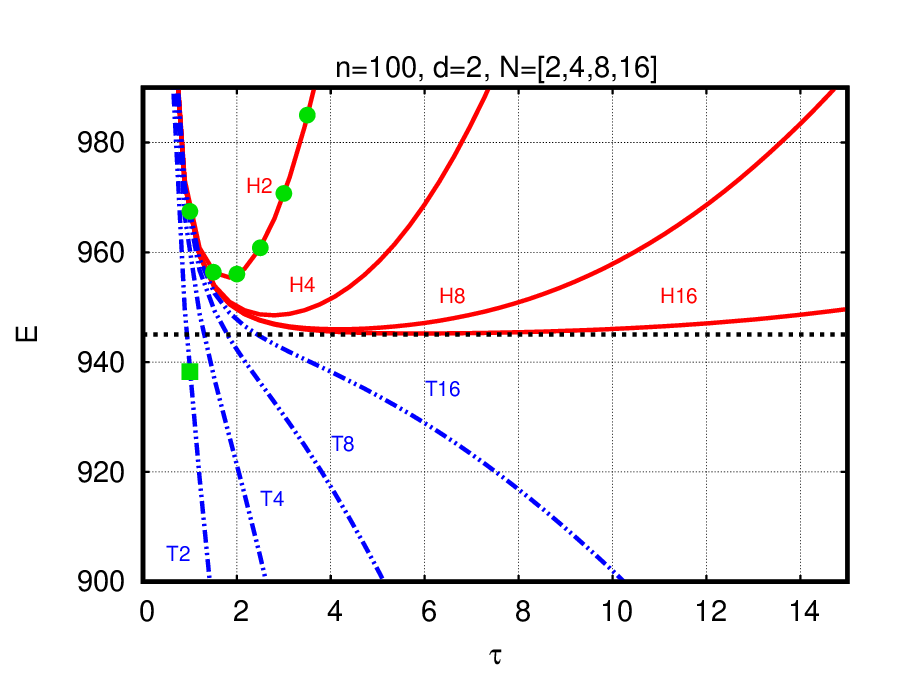}
    \caption{The convergence of various energies as functions of $\tau$, labeled by bead number $N=2,4,8,16$ for 100 non-interacting harmonic fermions.
    Solid red lines and dash-dotted blue lines denote Hamiltonian (H) and thermodynamic (T) energies, respectively. 
    The dotted black line is the exact ground state energy. Green symbols give direct PIMC simulation results for $N=2$ with disks and squares denoting Hamiltonian and thermodynamic energies, respectively.}
    \label{Fig:n100}
\end{figure}

For $d=2$, this difference scales as $E_F^2 - E_B^2 \sim n^{3/2}$, implying an explosive growth in the required numerical precision.
For example, an $n=100$ non-interacting harmonic fermion calculation, as shown in Fig.\ref{Fig:n100},
requires $\sim 2\times 10^4$ bits or $\sim 6{,}000$ decimal digits of precision. 

\begin{figure}[h]
    \includegraphics[width=\linewidth]{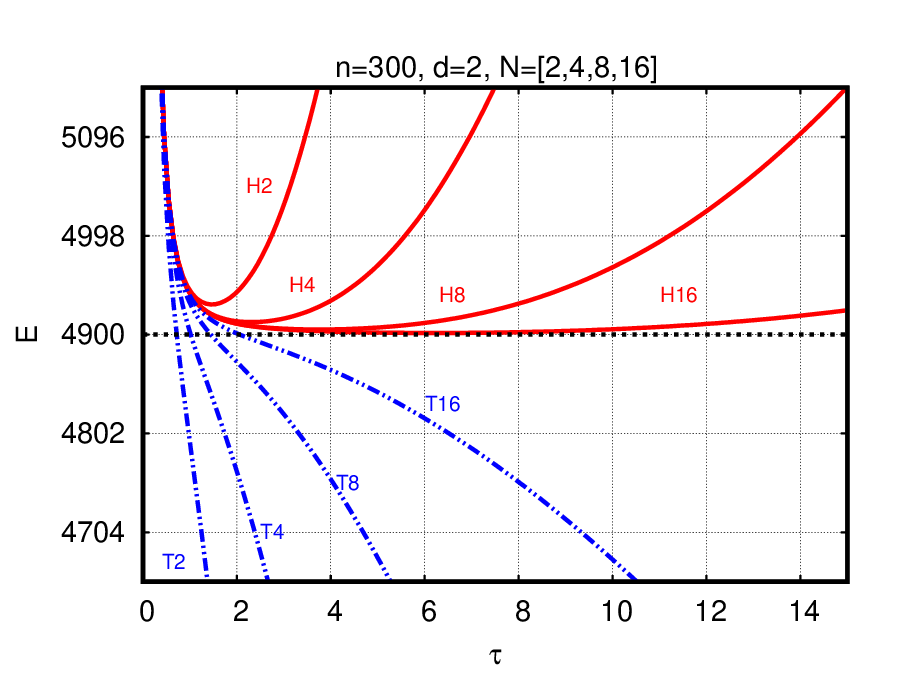}

    \caption{Similar plot to Fig.\ref{Fig:n100} for 300 non-interacting harmonic fermions.
     }
    \label{Fig:n300}
\end{figure}

For an $n=300$ free harmonic fermion calculation as shown in Fig.\ref{Fig:n300}, the required precision jumps to $\sim 10^5$ bits or $\sim 30{,}000$ decimal digits! 
As verified in Appendix~\ref{appx:precision},
the time complexity scales as $O(n^4\tau^{4/3})$, rendering the direct numerical evaluation of the recursive relation impractical for $n>300$.

In the following section, we show that this sign problem surprisingly can be completely bypassed for two-dimensional fermions through a subtraction-free algebraic reformulation in terms of $\lambda$-rings.

\section{Algebraic Relation and Closed-Form Partition Functions}
\label{appC}
The recursive relation (\ref{mainRecursion}) can be solved exactly for $d=1$ and $d=2$ by identifying it with the Newton identities, which leads to a natural $\lambda$-ring structure\cite{doi:10.1142/7664}. The corresponding dictionary is
\begin{center}
\begin{tabular}{c  c  c  c  c} 
 $\lambda$-ring &  & \qquad $Z_n$ & & symmetric functions \\ [0.5ex] 
 \hline\hline
 $\lambda^n(X)$ & \qquad $\leftrightarrow$ \qquad & \qquad $Z_n^{N,d}$ & \qquad $\leftrightarrow$ & $e_n$ \\
 $\psi^n(X)$ & \qquad $ \leftrightarrow$ \qquad & \qquad $z_n^{N,d}$ & \qquad $\leftrightarrow$ & $p_n$ \\
 $\lambda_t(X)$ & \qquad $ \leftrightarrow$ \qquad & \qquad $\Xi^N_d(t)$ & \qquad $\leftrightarrow$ & $E(t)$ \\
\end{tabular}
\end{center}
where $\Xi_d^N(t) = \sum_{k=0}^\infty Z_k^{N,d}t^k$ is the generating function of $Z_n^{N,d}$. If $t$ is identified with the fugacity ($t=e^{\tau \mu}$), then $\Xi_d^N$ is the grand-canonical partition function.

For the harmonic oscillator, this structure admits a concrete realization by taking $K = \mathbb Q[[b]]$, the ring of formal power series in $b$,
equipped with its canonical $\lambda$-ring structure $\Lambda(K) =1+K[[t]]^+$\cite{doi:10.1142/7664}. We factor the zero-point contribution $b^{nd/2}$ out of the recursion and restore it later. Under the identification $\psi^n(X) \leftrightarrow z_n^{N,d}$, one has the power series expansion
\begin{equation*}
    z_n^{N,d} =\psi^n(X) = \sum_{k\ge 0} \binom{k+d-1}{d-1}\, b^{nk},
\end{equation*}
giving $\psi^n(b^\alpha) = b^{n\alpha}$ and $X = (1-b)^{-d} = z_1^{N,d}$.

For $d=1$, the closed form for $Z_n^{N,1}$ follows from standard symmetric function identities\cite{NewtonIden}. Restoring the zero-point factor gives
\begin{equation}
    Z_n^{N,1} = \frac{b^{\frac{n^2}{2}}}{(b)_n}\,,
    \label{z1d}
\end{equation}
where $(b)_n = (b;b)_n =\prod_{k=1}^n(1 - b^k)$ is the q-Pochhammer symbol.

For $d = 2$, the dual Cauchy identity yields a closed form in terms of the difference of two Mahonian statistics, $\mathrm{maj}(\pi)-\mathrm{inv}(\pi)$ for permutations $\pi\in S_n$, where $\mathrm{maj}(\pi)$ is the major index (the sum of descent positions, i.e., indices $i$ where $\pi(i) > \pi(i+1)$) and $\mathrm{inv}(\pi)$ counts the number of inversions (pairs $i<j$ with $\pi(i)>\pi(j)$)\cite{Stanley_Fomin_1999}. Writing $X=\hat{X}\hat{X}$ with $\hat{X}=z^{N,1}_1$, one obtains $\Xi_2^N(t)=\lambda_t(\hat{X}\hat{X})$, with expansion\cite{macdonald1998symmetric}
\begin{equation*}
    \lambda_t(\hat{X}\hat{X}) =  \sum_{\lambda} s_\lambda(1,b,b^2,\ldots) s_{\lambda'}(1,b,b^2,\ldots)t^{|\lambda|},
\end{equation*}
where the sum is over all partitions $\lambda$,
$s_\lambda$ are Schur functions, and $\lambda'$ is the corresponding conjugate partition. 
Extracting coefficients, $Z_n^{N,2}(b) = [t^n]\lambda_t(\hat{X}\hat{X})$, gives
\begin{equation*}
    Z_n^{N,2}(b)  =\sum_{\lambda \vdash n} s_\lambda(1,b,b^2,\ldots) s_{\lambda'}(1,b,b^2,\ldots),
\end{equation*}
summing over all partitions of $n$ only.

Using the Standard Young's Tableau (SYT) expansion of Schur functions\cite[Cor.(7.21.3) \& Cor.(7.21.5)]{Stanley_Fomin_1999} and
$\mathrm{maj}(T^t)=\binom{n}{2}-\mathrm{maj}(T)$\cite{Huang_2020} yields a sum over tableau pairs. 
Under the Robinson–Schensted–Knuth (RSK) correspondence\cite[Eq.(7.23.1)]{Stanley_Fomin_1999}, the tableau pairs map to permutations $\pi\in S_n$ with the statistics mapping to $\mathrm{maj}(\pi)$ and $\mathrm{maj}(\pi^{-1})$. 
Applying Theorem~1 from Ref.\onlinecite{Foata1978MajorIA} and restoring the zero-point factor yields
\begin{equation}
Z^{N,2}_n(b)
=\frac{b^{\binom{n + 1}{2}}}{((b)_n)^2}
\sum_{\pi\in S_n} b^{\mathrm{maj}(\pi)-\mathrm{inv}(\pi)}.
\label{eq:Z2_closed}
\end{equation}
The permutation statistic in (\ref{eq:Z2_closed}) encodes the degeneracy structure of the $d=2$ spectrum, which is absent in $d=1$. For $d\ge 3$, a similar analysis would require the use of higher plethysms (e.g., $X = \hat{X}\hat{X}\hat{X}$).
We will defer such an analytical derivation to a future study. 

\section{Polynomial-Time Evaluation Schemes}
For $d=1$, the closed-form expression (\ref{z1d}) has been directly evaluated with $O(n)$ complexity\cite{NewtonIden}.

For $d=2$, the direct sum over $n!$ permutations in (\ref{eq:Z2_closed}) 
is impractical even for modest values of $n$. Fortunately,
Baxter and Zeilberger\cite{BaxterZeilberger_invmaj} have derived a
fast algorithm for computing 
\begin{equation*}
H_n(p,q) = \sum_{\pi\in S_n} p^{\mathrm{maj}(\pi)} q^{\mathrm{inv}(\pi)},
\end{equation*}
from which the required permutation sum in 
(\ref{eq:Z2_closed}) can be obtained as $H_n(b,1/b)$.
Furthermore, from Ref.\onlinecite{BaxterZeilberger_invmaj}, one can identify $H_n(b,1/b)=F(n+1,n+1)$,
where the auxiliary function $F(n,i)$ is given by a purely {\it additive} recursion
\begin{equation}
F(n,i) = b\,F(n,i+1) + b^{1-i}(1-b^{n-1})F(n-1,i),
\end{equation}
with boundary condition
\begin{equation}
F(n,n) = \sum_{j=1}^{n-1} F(n-1,j) \,,\quad F(1,1) = 1.
\end{equation}
The strictly positive terms in the above two equations
allow evaluation in the log-domain using log-sum-exp techniques, bypassing the instability from exponentially small contributions, yielding an $O(n^2)$ algorithm for evaluating $Z_n^{N,2}$. 

\begin{figure}[H]
    \includegraphics[width=\linewidth]{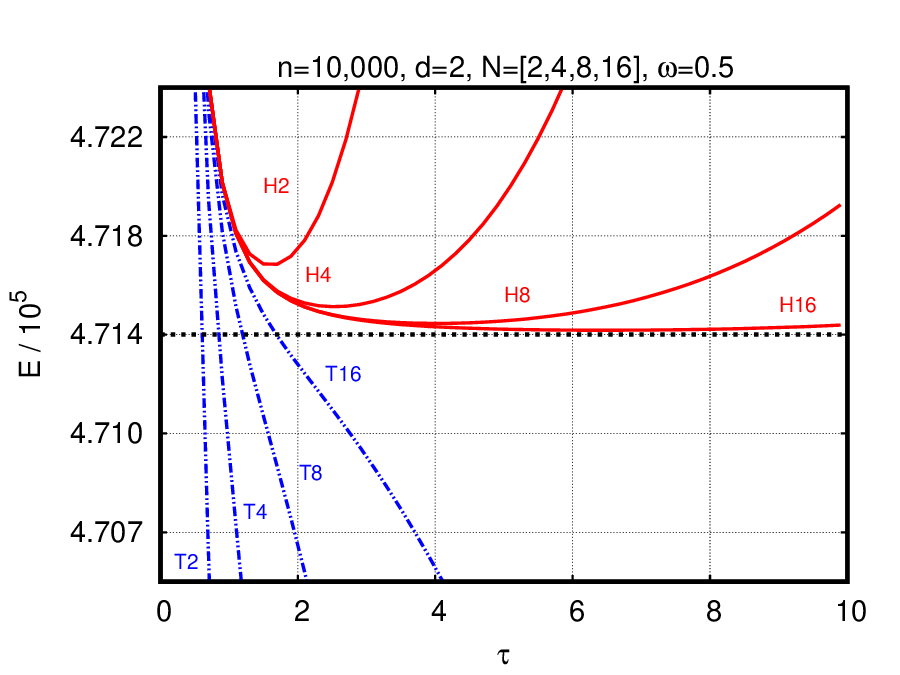}
    \caption{Similar plot to Fig.\ref{Fig:n100} for $n=10{,}000$ harmonic fermions with pairwise repulsive $\lambda<0$ harmonic interactions such that $\omega=0.5$.}
    \label{Fig:n10000}
    
    
    \includegraphics[width=\linewidth]{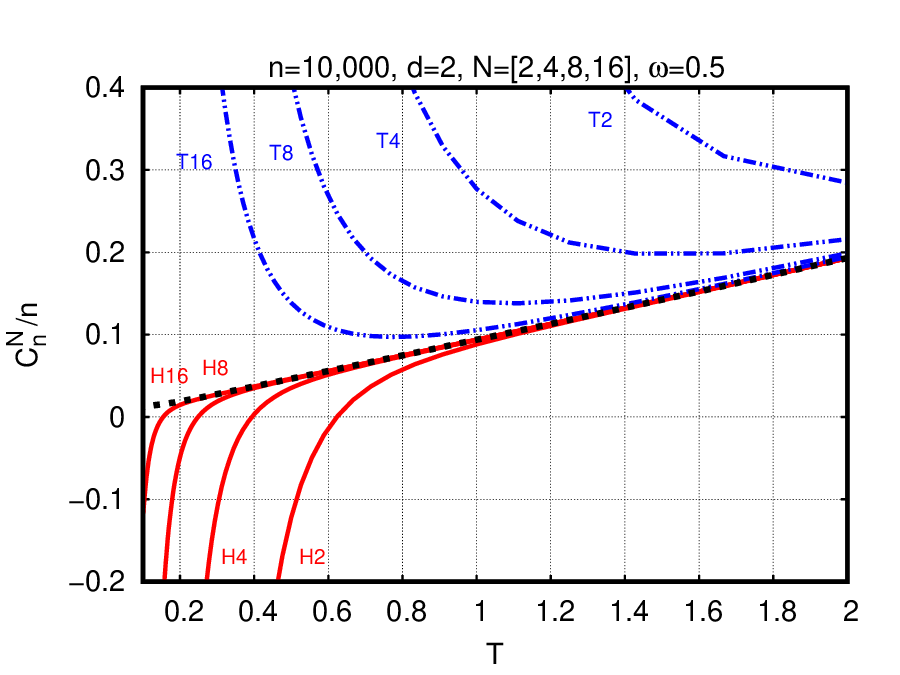}
    \caption{The specific heat per particle as in (\ref{cvn}) as functions of $T = 1/\tau$ labeled by bead number $N=2,4,8,16$ for $n=10{,}000$ harmonic fermions with pairwise repulsive harmonic interactions such that $\omega=0.5$. The dotted black line is the exact continuum result.}
    \label{cvn105}
\end{figure}

Since we have the $N$-bead partition function, one can calculate not only the energy in Fig.\ref{Fig:n10000} but also the specific heat 
\begin{equation}
  C_n^N = \frac{\partial E^{N}_n}{\partial T}
  \label{cvn}
\end{equation}
by finite differences as shown in Fig.\ref{cvn105}. The specific heat per particle correctly approaches the value of 2 in the high-temperature limit.
\section{Conclusion and Future Directions}

In this work, we have shown that after analytically evaluating the discrete path integrals for 
two-dimensional harmonic fermions, the resulting canceling recursion relation, through algebraic restructuring, 
can be evaluated with a purely additive scheme, leading to a stable $O(n^2)$-complexity algorithm for computing the original harmonic PIMC partition function with a well-documented sign problem. This provides, to our knowledge, the first subclass of 
explicitly fermionic sign problems, with no possible bosonic mapping, but solvable in polynomial time.
This is a clear counterexample to the prevailing view\cite{tro05} that {\it all} fermionic PIMC sign problems that are not NP-hard require an essentially bosonic description. 

Moreover, the original sign problem in PIMC is conventionally defined\cite{tro05} as
$\langle s \rangle=Z_F/Z_{|F|}$,
where $Z_F$ is the exact fermion partition function and $Z_{|F|}$ is the partition function evaluated ignoring the sign of the integrand. It then follows that $Z_{|F|}\ge Z_F$ and $\langle s \rangle\le 1$.
Since at large $\tau$, the boson partition function $Z_B\sim\exp(-E_B\tau)$
is much, much greater than $Z_F\sim\exp(-E_F\tau)$ or $Z_{|F|}$, the sign problem defined by
$\langle s' \rangle=Z_F/Z_B$ is then much, much worse than $\langle s\rangle$.
But this is exactly the sign problem we initially tried to overcome via high precision, since
$\langle s'\rangle\approx {\rm e}^{-B_2}$.
Therefore, our algebraic restructuring has solved an even more 
virulent form of the original harmonic fermion sign problem, in quadratic time. This strongly suggests that the original harmonic fermion PIMC sign problem, prior to exact integration, is also not NP-hard.

For a general potential $V(r)$ one can approximate
\begin{equation*}
e^{-\tau V(r)} \approx \sum_{k=1}^{p} \alpha_k(\tau)\, e^{-\beta_k(\tau)|r - v_k(\tau)|^2},
\end{equation*}
by a finite sum of displaced Gaussian functions\cite{heh69,Sandberg2001GaussianRBF}.
According to Ref.\onlinecite{heh69}, no more than six Gaussians are needed to approximate most Slater wave functions.
Thus the present method may also be applicable to 
a wider class of fermionic systems.

\begin{acknowledgments}
This work was supported by the internal funding of the Department of Physics and Astronomy at Texas A\&M University. Portion of this research was conducted using advanced computational resources provided by Texas A\&M High Performance Research Computing.
\end{acknowledgments}
\medskip

A.C. wrote the initial draft and provided the mathematical method of solution. J.V. carried out the numerical computations and prepared the graphical presentations. S.C. contributed to the final draft and provided the original impetus for this work.

\medskip
\noindent{\bf Data availability ---} The source code and simulation data that support the findings of this study are openly available on GitHub and are permanently archived in Zenodo at Ref.\onlinecite{github_code}.

\bibliographystyle{apsrev4-2}
\bibliography{ref} 


\appendix

\section{Precision requirements for evaluating the recursion relation}
\label{appx:precision}

In order to avoid the numerical instability resulting from alternate-sign cancellations in (\ref{mainRecursion}), the recursion relation must be computed with high precision. The magnitude of the largest term in (\ref{mainRecursion}) 
is either $z_n^d$ or $z^d_{n-1}Z^d_1=z^d_{n-1}z^d_1$, 
both of which are less than
the $n$-boson partition function $Z_B^d = (z_1^d)^n$. Starting from these large terms, 
we must, through cancellations, arrive at the small $n$-fermion partition function $Z_F^d$. 
Therefore, in order for $Z_F^d$ to be correct to at least one digit (and hence stable), 
we require the evaluation of these large terms, which are at most approximately $Z^d_B$, 
be correct to the same order of magnitude as $Z_F^d$.

\begin{figure}[t]
    \includegraphics[width=\linewidth]{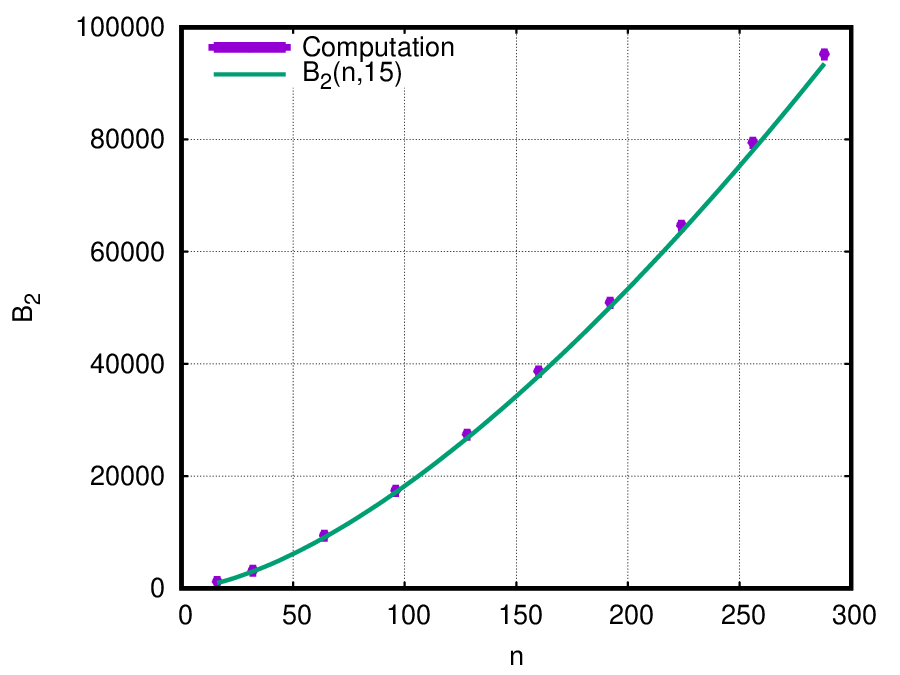}
    \caption{Empirically determined minimum number of bits determined by the method described in the text (purple dots), compared to analytic estimation (\ref{bits2d}) (green solid line).}
    \label{Fig:Scaling1}
\end{figure}

The required bit precision $B_d(n,\tau)$ is therefore approximately
the difference between the magnitudes of these two quantities in base $2$,
\begin{equation}
    B_d(n,\tau) \approx \log_2(Z_B^d) - \log_2(Z_F^d).
\end{equation}
In the continuum limit, $\ln Z \approx -\tau E$, giving
\begin{equation}
    B_d(n,\tau) \approx \frac{\tau}{\ln 2}(E_F^d - E_B^d).
    \label{eq:Bd}
\end{equation}
For $d=2$, with $E_B^2=n$ and 
$E_F^2 \approx \frac{2\sqrt{2}}{3}n^{3/2}$, this gives
\begin{equation}
    B_2(n,\tau) = \frac{2\sqrt{2}}{3\ln 2}
    \!\left(n^{3/2}-\tfrac{3\sqrt{2}}{4}n\right)\tau,
    \label{bits2d}
\end{equation}
with $n=300$ requiring $B_2\!\sim\!10^5$ bits of precision. To validate (\ref{bits2d}), we computed $Z_F^2$ for several values of $n$ using the recursion relation (\ref{mainRecursion}) with 
arbitrary-precision arithmetic (ArbNumerics/GMP). For each $n$, a reference value was  established using $B_2(n,\tau) + 2000$ bits of precision, guaranteeing $\sim 600$ correct decimal digits. We then increased the working precision from 500 bits in increments of 250 bits until the maximum relative error in $Z_F^2$, evaluated for 20 values of $\tau\in[5,15]$, was at most $10^{-8}$. This resulting empirical minimum precision, shown in Fig.\ref{Fig:Scaling1} agrees closely with (\ref{bits2d}).

\begin{figure}[H]
    \includegraphics[width=\linewidth]{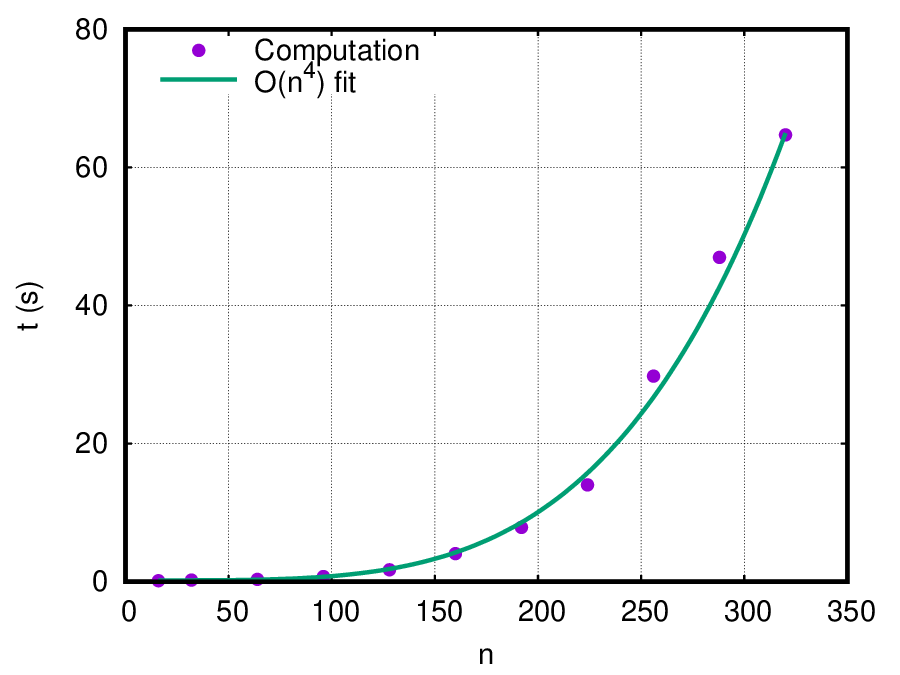}
    \caption{Computation time vs. $n$ comparison between (\ref{Fig:Scaling3}) and the direct recursion algorithm with $N=200$, $d=2$ and $\tau=15$.}
    \label{Fig:Scaling2}
\end{figure}
In the range of $10^4 -10^5$ bits, the GMP multiplication algorithm mainly operates in the Toom-8.5 regime with time complexity $O(B^{4/3}_d)$\cite{gmp_manual_fft}. Since the recursion (\ref{mainRecursion}) requires $O(n^2)$ such operations, and the $z_k^d$ values can be precomputed in $O(n \cdot B_d^{4/3})$, the dominant cost is
\begin{equation}
\label{Fig:Scaling3}
    t \sim O\!\left(n^2 \cdot B_d(n,\tau)^{4/3}\right).
\end{equation}
For $d=2$, substituting $B_2 \sim n^{3/2}\tau$ yields $O(n^4\tau^{4/3})$, confirmed in Fig.\ref{Fig:Scaling2}, making direct recursive evaluation 
infeasible beyond $n\sim 300$.

\section{PIMC comparison to semi-classical approximation}

\begin{figure}[H]
    \vspace{2em}
    \includegraphics[width=\linewidth]{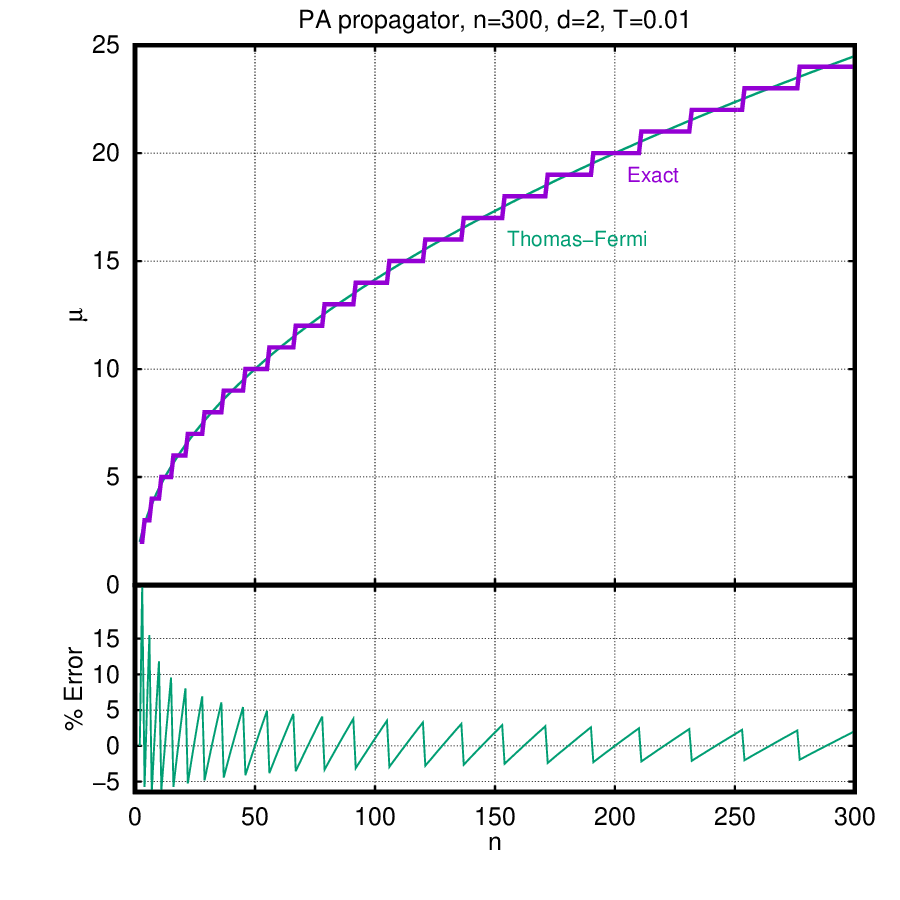}
    \vspace{-2em}
    \caption{Comparison of the chemical potential computed via the algebraic restructuring method and the Thomas-Fermi (TF) semiclassical approximation in the continuum limit with $d=2$ at $T=1/\tau = 0.01$. The restructuring method preserves the discrete shell-filling plateaus that the TF approximation smoothes over.}
    \label{Fig:Muvsn}
\end{figure}

To illustrate the exactness of the algebraic formulation, Fig.\ref{Fig:Muvsn} compares the chemical potential $\mu$ obtained from the algebraic restructured partition function with the Thomas-Fermi semiclassical approximation. The Baxter-Zeilberger evaluation explicitly preserves the discrete shell-filling structure of the two-dimensional system, which manifests as distinct plateaus in the chemical potential. In contrast, the semiclassical approximation is incapable of reproducing these quantum shell-structure effects.

This comparison demonstrates that the algebraic reformulation exactly 
preserves the original quantum statistics of the finite $n$-fermion system,
which is not directly computable via PIMC.

\end{document}